\documentclass[twocolumn,letterpaper]{ieeeconf}                                                          

\IEEEoverridecommandlockouts
\usepackage{amssymb,amsthm,amsmath}
\usepackage{graphicx,subfigure,enumerate,color}
\usepackage{bbold,cite}
\usepackage[hide links]{hyperref}
\usepackage{algorithm}
\usepackage[no end]{algorithmic}

\newcommand{\Rpp}{\mathbb{R}_{> 0}}

\theoremstyle{theorem}

\theoremstyle{remark}

\renewcommand{\baselinestretch}{0.971}

\title{A Modified SIR Model for the COVID-19 Contagion in Italy}

\author{Giuseppe C. Calafiore, Carlo Novara\thanks{G.C. Calafiore and C. Novara are with
Dipartimento di Elettronica e Telecomunicazioni, Politecnico di Torino, Corso Duca degli Abruzzi 24, 10129 Torino, Italy
(e-mail: \{giuseppe.calafiore, carlo.novara\}@polito.it).
G.C. Calafiore is also with Istituto di Elettronica e di Ingegneria dell'informazione e delle Telecomunicazioni, Consiglio Nazionale delle Ricerche (IEIIT-CNR), c/o Politecnico di Torino, Corso Duca degli Abruzzi 24, 10129 Torino, Italy.} and Corrado Possieri\thanks{C. Possieri is with Istituto di Analisi dei Sistemi ed Informatica ``A. Ruberti'', Consiglio Nazionale delle Ricerche (IASI-CNR), via dei Taurini 19, 00185 Roma, Italy (e-mail: corrado.possieri@iasi.cnr.it). This work has been completed on March 30, 2020.}
}

\begin{document}
\maketitle

\begin{abstract}
The purpose of this work is to give  a contribution to the understanding of the COVID-19 contagion in Italy. 
To this end, we developed a modified Susceptible-Infected-Recovered (SIR) model for the contagion, and we used official data of the pandemic up to March 30th, 2020 for identifying the parameters of this model. The non standard part of our approach resides in the fact that we considered as  model parameters also the initial number of susceptible individuals, as well as the proportionality factor relating the detected number of positives with the actual (and unknown) number of infected individuals. Identifying the contagion, recovery and death rates as well as the mentioned parameters amounts to a non-convex identification problem that we solved by means of a two-dimensional grid search in the outer loop, with a standard weighted least-squares optimization problem as the inner step.
\end{abstract}

\section{Introduction}
Mathematical models can offer a precious tool to public health authorities for the control of epidemics, potentially
contributing to  significant reductions in
the number of infected people and deaths.
 Indeed, mathematical models
can be used for obtaining short and long-term predictions, which in turn
may enable decision makers  optimize possible control strategies,
such as containment measures, lockdowns  and vaccination campaigns. Models can also
be crucial in a number of other tasks, such as estimation of transmission
parameters, understanding of contagion mechanisms, simulation of
different epidemic scenarios, and test of various hypotheses.

Several kind of models have been proposed for describing the time evolution
of epidemics, among which we distinguish two main groups: collective
models and networked models. Collective models are characterized by
a small number of parameters and describe the epidemic spread in a population
using a limited  number of collective variables. They include generalized
growth models \cite{Chowell2017}, logistic models \cite{SIR1927},
Richards models \cite{Richards1959}, Generalized Richards models
\cite{Chowell2017}, sub-epidemics wave models \cite{Chowell2019},
Susceptible-Infected-Recovered (SIR) models \cite{SIR1927,bailey1975mathematical},
and Susceptible-Exposed-Infectious-Removed (SEIR) models \cite{Chowell2017}.
SIR, SEIR and other similar models belong to the class of the so-called
compartmental models \cite{Brauer2017,Chowell2017}. Networked models
typically treat a population as a network of interacting individuals
and the contagion process is described at the level of each individual,
see, e.g., \cite{Keeling2005,Nadini2018,Nowzari2017,Pastor-Satorras2015,PastorePiontti2014,Pellis2015,Wertheim2014}.
These models clearly provide a more detailed description of the epidemic
spread than collective models but their identification is significantly
harder. A first reason is that they are usually characterized by a
high number of parameters and variables. A second reason, perhaps more
relevant, is that the network topology is unknown in most real situations
and its identification is an extremely hard task. In this paper, we
focus on collective models since, thanks to their relative simplicity,
they can be more suitable for non-expert operators and public health
authorities, and they can provide simple but reliable models, even under scarcity of
data.

Collective models are typically written in the form of differential
equations or discrete-time difference equations, and are characterized
by a set of parameters that are not known a-priori and have to be
identified from data. 
However, the identification of such parameters
raises several practical issues, as discussed next.  
An important variable in many epidemic models is the number
of individuals that are infected at a given time. However, in a real
epidemic scenario, only the number of infected individuals that have
been detected as ``positive'' is available, while the actual  number of infected people
remains unknown. A common assumption made in the literature is that the
observed cases are the actual ones. Clearly, this assumption is unrealistic
and may lead to wrong epidemiological interpretations/conclusions. 
Other issues stem from the fact that identification of epidemic models requires in many cases to deal with non-convex optimization problems.
Indeed, a key feature of an epidemic
model is to provide reliable results in long-term predictions, in
order to allow analysis/comparison of different scenarios and design
of suitable control strategies. Hence, identification has to be performed
with the objective of  minimizing the model multi-step prediction error. This typically
requires solving a non-convex optimization problem, even when the
model is linear in the parameters, with the ensuing relevant risk of 
being trapped in poorly-performing local solutions.
Furthermore, the initial values of some model
variables have often to be identified, in addition to the model parameters, and 
this also requires solution of a non-convex optimization problem.

In this paper, we propose a variant of the SIR model, developed
in order to describe the actual number of infected individuals. As discussed above,
this quantity is important from the epidemiological standpoint.
The second contribution consists in a model identification and prediction framework that allowed
us to overcome the mentioned problems in the modeling of the infection evolution  of the present  COVID-19 pandemics.
The model identification approach
is based on a simple yet practically effective scheme: a  model structure
is assumed, characterized by a set of parameters to be identified.
A grid is defined for those few (two, in the actual model considered here) parameters on which the model has a nonlinear
dependence (with some abuse of terminology, these are called the nonlinear
parameters). For each point of this grid, the other parameters are
identified via convex optimization. Finally, the optimal parameter
estimate is chosen as one minimizing a suitable objective function
over the grid. This approach is particularly suitable for epidemic
collective models, which typically feature a low number of nonlinear
parameters. Clearly, when the number of such parameters is large,
the approach becomes computationally unfeasible. 

In general, this approach is expected to provide reliable parameter
estimates. However, the resulting model may be not extremely precise
in long-term predictions, since convex optimization allows minimization
of the one-step prediction error, but not minimization of multi-step
prediction errors. To overcome this issue, we employed a novel long-term
prediction algorithm, based on a weighted average of the multi-step
predictions performed by starting the simulation at all the available initial conditions. 
The weighted
average allows a reduction of noise and error effects, possibly yielding significant
improvements in the long-term prediction accuracy.


A real-data case study is presented, concerned with the current COVID-19 epidemic in Italy.

\section{SIRD model for COVID-19 contagion}
We consider a geographical region, assumed as isolated from other regions, and within such region we define:
\begin{itemize}
\item $S(t)$: the number of individuals \emph{susceptible} of contracting the infection at time $t$;
\item $I(t)$: the number of \emph{infected} individuals that are active at time $t$;
\item $R(t)$: the cumulative number of individuals that \emph{recovered} from the disease up to  time $t$;
\item $D(t)$: the cumulative  number of individuals that \emph{deceased} due to the disease, up to time $t$.
\end{itemize}
We thus seek to describe
approximately 
the dynamics of the COVID-19 infection via the following discrete-time  version of the Kermack-McKendrick  equations, as given in \cite{bailey1975mathematical}, 
so to account for the number of deaths due to the infection:
\begin{subequations}
\label{eq:SIRD}%
\begin{align}
S(t+1) & = S(t) - \beta\,\frac{S(t)\,I(t)}{S(t)+I(t)},\\
I(t+1) & = I(t) + \beta\,\frac{S(t)\,I(t)}{S(t)+I(t)}-\gamma \,I(t)-\nu\,I(t),\\
R(t+1) & = R(t) + \gamma\,I(t),\\
D(t+1) & = D(t) + \nu\,I(t),
\end{align}
\end{subequations}
with initial conditions $S(t_0)=S_0>0$, $I(t_0)=I_0>0$, $R(t_0)=R_0\geq 0$ and $D(t_0)=D_0 \geq 0$,
where $\beta\in\Rpp$ is the \emph{infection rate}, $\gamma\in\Rpp$ is the \emph{recovery rate}, and $\nu\in\Rpp$ is the \emph{mortality rate}. 
Time $t=0,1,\ldots$ is here expressed in days. 
Equations \eqref{eq:SIRD} are a discrete-time version of the classical Susceptible-Infected-Recovered (SIR) model.
The underlying hypotheses in this model are that the recovered subjects are no longer susceptible of infection (an hypothesis which is apparently not yet proved, or disproved, for COVID-19), and that the number of deaths due to 
other reasons (different from the disease under consideration) are neglected by the model. 
Further, each region is assumed to be isolated from other regions, which could be a reasonable assumption if
containment measures are enforced.

Model~\eqref{eq:SIRD} assumes that the value $I(t)$ is the actual number of infected individuals. Nonetheless, in practice, observations of the process only permit to detect a portion $\tilde{I}(t)$ of infected individuals, since some of them may be asymptomatic \cite{mizumoto2020estimating}.
We assume that such a number is an (unknown) fraction of the actual number $I(t)$, that is 
\begin{equation}\label{eq:detInf}
I(t) = \alpha \tilde I(t), \quad \text{for some } \alpha \geq 1.
\end{equation}

By plugging~\eqref{eq:detInf} into~\eqref{eq:SIRD},
we obtain the following model 
\begin{subequations}
\label{eq:SIRDm}%
\begin{align}
\tilde{S}(t+1) & = \tilde{S}(t) - \beta\,\frac{\tilde{S}(t)\,\tilde{I}(t)}{\tilde{S}(t)+\tilde{I}(t)},\\
\tilde{I}(t+1) & = \tilde{I}(t) + \beta\,\frac{\tilde{S}(t)\,\tilde{I}(t)}{\tilde{S}(t)+\tilde{I}(t)}-\gamma \,\tilde{I}(t)-\nu\,\tilde{I}(t),\\
\tilde{R}(t+1) & = \tilde{R}(t) + \gamma\,\tilde{I}(t),\\
D(t+1) & = D(t) + \alpha\,\nu\, \tilde I(t),
\end{align}
\end{subequations}
where $\tilde{S}(t):=\frac{1}{\alpha}\,S(t)$ denotes the weighted susceptible individuals at time $t$, and
$\tilde{R}(t):=\frac{1}{\alpha}\,R(t)$ denotes the detected recovered individuals at time $t$.
In the following, equations \eqref{eq:SIRDm} will be referred to as the SIRD model.

As in its continuous-time counterpart \cite{gleissner1988spread}, the dynamics of systems~\eqref{eq:SIRD} and~\eqref{eq:SIRDm} are highly dependent on the initial conditions $S(t_0)=\alpha\,\tilde{S}(t_0)$ and $I(t_0):=\alpha\,\tilde{I}(t_0)$, which determine both the amplitude and the time location of the peak in the number of infected individuals.
Unfortunately, the datum $S(t_0)$ is not available to the modeler (notice that taking $S(t_0)$ equal to the total population
of the region of interest may be a gross over-estimation 
of the initial number of susceptible individuals, since part of the population may be inherently immune or non affected by the contagion), thus rendering the problem of making predictions via~\eqref{eq:SIRD} and~\eqref{eq:SIRDm} rather challenging.
The main objective of this paper is then to estimate the parameters $S(t_0)$, $\alpha$, $\beta$, $\gamma$, and $\nu$ of the model
from available data, 
so to accurately predict the behavior of the COVID-19 spread in Italy.

\section{Model identification}
In this section, we detail the procedure that has been used to identify the parameters 
$S(t_0)$, $\alpha$, $\beta$, $\gamma$, and $\nu$ of the SIRD model in~\eqref{eq:SIRDm}.
The data that have been used to carry out the identification are the official
data from Italian {\em Dipartimento della Protezione Civile}, available at
\begin{center}
\url{https://github.com/pcm-dpc/COVID-19}, 
\end{center}
and are constituted by the numbers $\tilde{I}(t)$, $\tilde{R}(t)$, and $D(t)$, where the discrete time
represent the number of days from the start of the epidemy, over a time window starting from
February 24th, 2020 and ending March 30th, 2020.

\begin{figure}[htb!]
\centering
\includegraphics[width=0.9\columnwidth]{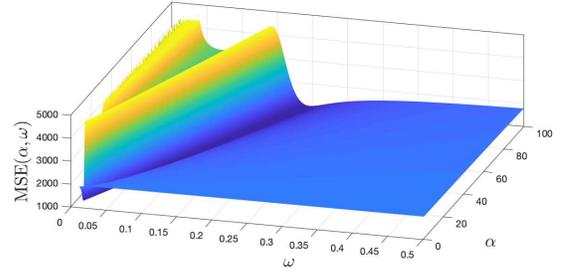}
\caption{Value of $\mathrm{MSE}(\alpha,\omega)$ considering Italy as a single region.\label{fig:surface}}
\end{figure}

In order to correctly represent the number of susceptible individuals, we introduced an additional parameter
$\omega\in[0,1]$ such that $S(t_0)=\omega\,P$, where $P$ is the total population in the region under examination,
and we defined 
\begin{equation}\label{eq:tildeS}
\tilde{S}(t)=\frac{\omega}{\alpha}\,P - \tilde{I}(t) - \tilde{R}(t)-\tilde{D}(t).
\end{equation}
Hence, for fixed values of $\omega$ and $\alpha$, the model~\eqref{eq:SIRDm} can be expressed in regression form
\begin{equation*}
\Delta(t) :=
 \left[ \begin{array}{c}
 \tilde I(t+1) - \tilde I(t) \\
  \tilde R(t+1) - \tilde R(t) \\
 D(t+1) - D(t)
  \end{array}\right]=\Phi_{\omega,\alpha}(t)
    \left[ \begin{array}{c} 
    \beta \\
    \gamma \\
     \tilde \nu
    \end{array}\right],
\end{equation*}
where $\tilde{\nu}:=\alpha\,\nu$ and
\begin{equation*}
\Phi_{\omega,\alpha}(t):= \left[\begin{array}{ccc}
    \frac{\tilde S(t) \tilde I(t)}{\tilde S(t) + \tilde I(t)} & -\tilde I(t) & -\tfrac{1}{\alpha}\,\tilde I(t)  \\
   0 & \tilde I(t)  & 0 \\
   0 & 0 & \tilde I(t) 
 \end{array}\right].
\end{equation*}
By stacking the weighted vectors $\Delta(t)$ and the matrices $\Phi_{\omega,\alpha}(t)$ over the available time window we obtain the matrices
\begin{subequations}
 \label{eq_regression}%
\begin{align}
\overline{\Delta} & = \left[\begin{array}{c}
\rho^{\Theta-t_0} \Delta(t_0)\\
\rho^{\Theta-t_0-1} \Delta(t_0+1)\\
\vdots\\
 \Delta(\Theta)
\end{array}\right],\\
\overline{\Phi}_{\omega,\alpha} & = \left[\begin{array}{c}
\rho^{\Theta-t_0} {\Phi}_{\omega,\alpha}(t_0)\\
\rho^{\Theta-t_0-1} {\Phi}_{\omega,\alpha}(t_0+1)\\
\vdots\\
 {\Phi}_{\omega,\alpha}(\Theta)
\end{array}\right],
\end{align}
\end{subequations}
where $\rho\in(0,1)$ is an exponential decay weighting parameter, used to give more relevance to most recent data,
and $\Theta$ is the length of the time window. 
Then, the parameters $\beta$, $\gamma$, and $\tilde{\nu}$ can be estimated by solving
the mean square optimization problem 
\begin{equation}\label{eq:means}
\mathrm{MSE}(\alpha,\omega):=\min_{\beta,\gamma,\tilde{\nu}}\left\Vert \overline{\Delta} -\overline{\Phi}_{\omega,\alpha} \left[ \begin{array}{c} 
 \beta \\
 \gamma \\
  \tilde \nu
 \end{array}\right]\right\Vert_2^2
 \end{equation}
for given $\alpha$ and $\omega$.
The optimal solution to this problem is 
\begin{equation}\label{eq:MS}
\left[ \begin{array}{c} 
 \beta \\
 \gamma \\
  \tilde \nu
 \end{array}\right] = \overline{\Phi}_{\omega,\alpha}^\dagger \,\overline{\Delta},
\end{equation}
where $\overline{\Phi}_{\omega,\alpha}^\dagger$ denotes the Moore-Penrose pseudo-inverse 
of matrix $\overline{\Phi}_{\omega,\alpha}$.
It is worth pointing out that while the optimization problem~\eqref{eq:means}
is convex and hence readily solvable by convex optimization methods \cite{boyd2004convex}, the problem
\begin{equation}\label{eq:overallProb}
\min_{\alpha,\omega,\beta,\gamma,\tilde{\nu}}\left\Vert \overline{\Delta} -\overline{\Phi}_{\omega,\alpha} \left[ \begin{array}{c} 
 \beta \\
 \gamma \\
  \tilde \nu
 \end{array}\right]\right\Vert_2^2
 \end{equation}
need not be convex. As an example, Figure~\ref{fig:surface}
depicts the value $\mathrm{MSE}(\alpha,\omega)$
for $\alpha\in[1,100]$, $\omega\in[0,1]$ and $\rho=0.9$
considering all the Italian territory as a single region.
As shown in this figure, the function $\mathrm{MSE}(\alpha,\omega)$ is not convex, thus
making the problem of computing the solution to~\eqref{eq:overallProb} rather challenging.
Nonetheless, the parameters $\alpha$, $\beta$, $\gamma$, $\tilde \nu$, and $\omega$ 
can be determined by using the following Algorithm~\ref{alg:tune},
which computes the model parameters that better fit the data
by gridding the variables $\alpha$, $\omega$, using~\eqref{eq:MS}
to determine $\mathrm{MSE}(\alpha,\omega)$, and solving $\min_{\alpha,\omega}\mathrm{MSE}(\alpha,\omega)$.

\begin{tiny}
\begin{algorithm}[htb!]
\caption{Tuning of the model parameters\label{alg:tune}}
\begin{algorithmic}[1]
\REQUIRE{data $\tilde{I}(t)$, $\tilde{R}(t)$, and $D(t)$, maximum value $\overline{\alpha}$ of $\alpha$,
weighting parameter $\rho$,	and total population $P$}
\ENSURE{parameters of the model~\eqref{eq:SIRDm}}
\STATE gird uniformly the planar region $[0,1]\times[1,\overline{\alpha}]$
\STATE $e\leftarrow + \infty$
\FOR{ each value $(\omega,\alpha)$ in the grid}
	\STATE define $\tilde{S}$ as in~\eqref{eq:tildeS}
	\STATE compute the matrices $\overline{\Delta}$ and $\overline{\Phi}_{\omega,\alpha}$ as in~\eqref{eq_regression}
	\STATE determine the parameters  $\beta$, $\gamma$, and $\tilde \nu$ as in~\eqref{eq:MS}
	\IF{$\Vert \overline{\Delta} -\overline{\Phi}_{\omega,\alpha} [ \begin{array}{ccc} 
	 \beta &
	 \gamma &
	  \tilde \nu
	 \end{array}]^\top\Vert_2^2<e$}
		\STATE $e\leftarrow \Vert \overline{\Delta} -\overline{\Phi}_{\omega,\alpha} [ \begin{array}{ccc} 
			 \beta &
			 \gamma &
			  \tilde \nu
			 \end{array}]^\top\Vert_2^2$
		\STATE $\omega^\star \leftarrow \omega$ and  $\alpha^\star \leftarrow \alpha$
		\STATE $\beta^\star \leftarrow \beta$,  $\gamma^\star \leftarrow \gamma$, and $\tilde{\nu}^\star\leftarrow \tilde{\nu}$
	\ENDIF
\ENDFOR 
\RETURN $\omega^\star$, $\alpha^\star$, $\beta^\star$, $\gamma^\star$, and $\tilde{\nu}^\star$
\end{algorithmic}
\end{algorithm}
\end{tiny}

Table~\ref{tab:nation} reports the values of the parameters obtained 
using Algorithm~\ref{alg:tune} considering either each region disjointedly
or all the Italian cases of COVID-19 with $\overline{\alpha}=100$ and $\rho=0.9$.
%

\begin{table*}[htb]
\centering
\caption{Model parameters in Italy and in its regions\label{tab:nation}}
{\renewcommand{\arraystretch}{1.2}
\setlength{\tabcolsep}{16pt}
\begin{tabular}{l|ccccc}
Region & $\alpha$ & $\beta$ & $\gamma$ & $\tilde \nu$ & $\omega$\\
\hline
Abruzzo &   81.9764 &  0.254559 & 0.0102637 & 0.0112523 &       0.1874\\
Basilicata &     91.741 &   0.250907 & 0.00302937 & 0.00467164 &       0.0923\\
Calabria &     83.423 &   0.201084 & 0.00547199 & 0.00792437 &       0.0832\\
Campania &    54.7853 &   0.142671 & 0.00531024 & 0.00900027 &       0.157\\
Emilia Romagna &   60.3835 &   0.19317 & 0.0117399 & 0.0120007 &  0.222727\\
Friuli-Venezia Giulia &    62.7827 &   0.239275 &  0.0255812 & 0.00826425 &  0.0863636\\
Lazio &    84.3756 &    0.22341 &  0.0137916 & 0.00655099 &  0.0545455\\
Liguria &   26.7945 &  0.238016 & 0.0199404 & 0.0161546 & 0.0636364\\
Lombardia &   17.9974 &  0.189301 & 0.0307642 & 0.0208288 & 0.0863636\\
Marche &     25.9947 &    0.196325 & 0.000527925 &   0.0112068 &   0.0681818\\
Molise &   79.5772 &  0.197276 & 0.0167297 &  0.006787 &      0.352\\
Piemonte &    33.1924 &   0.231923 & 0.00606022 &  0.0104308 &  0.0772727\\
Puglia &    85.9751 &   0.211897 &  0.0029805 & 0.00664412 &       0.152\\
Sardegna &     24.518 &   0.213762 &  0.0100864 & 0.00538705 &       0.250\\
Sicilia &     43.672 &   0.195245 &  0.0112913 & 0.00831225 &       0.0512\\
Toscana &    41.1898 &   0.186643 & 0.00380713 & 0.00641778 &  0.0681818\\
Trentino-Alto Adige/S\"udtirol &   17.1976 &  0.213756 & 0.0170006 & 0.0104204 & 0.0590909\\
Umbria &    72.3795 &   0.347926 &  0.0311456 & 0.00387433 &  0.0863636\\
Valle d'Aosta &    10.7997 &    0.29359 & 0.00565177 &  0.0112402 &       0.0532\\
Veneto &    22.7958 &    0.19047 & 0.00938741 & 0.00509062 &       0.05\\[1ex]
\hline
Italy & 63.135 &0.21542 &0.017129 &0.011832 & 0.12384
\end{tabular}
}
\end{table*}

\section{Model Predictions }

\begin{figure*}[htb!]
\centering
\includegraphics[width=0.95\textwidth]{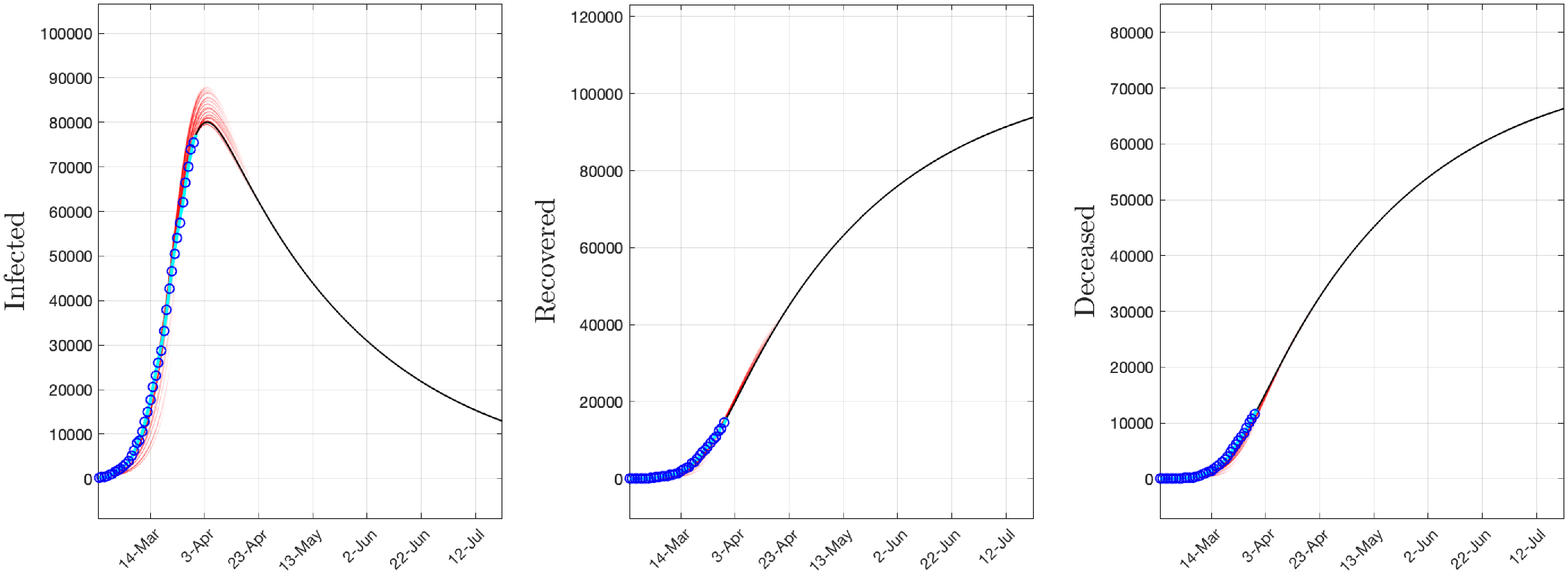}
\includegraphics[width=\columnwidth]{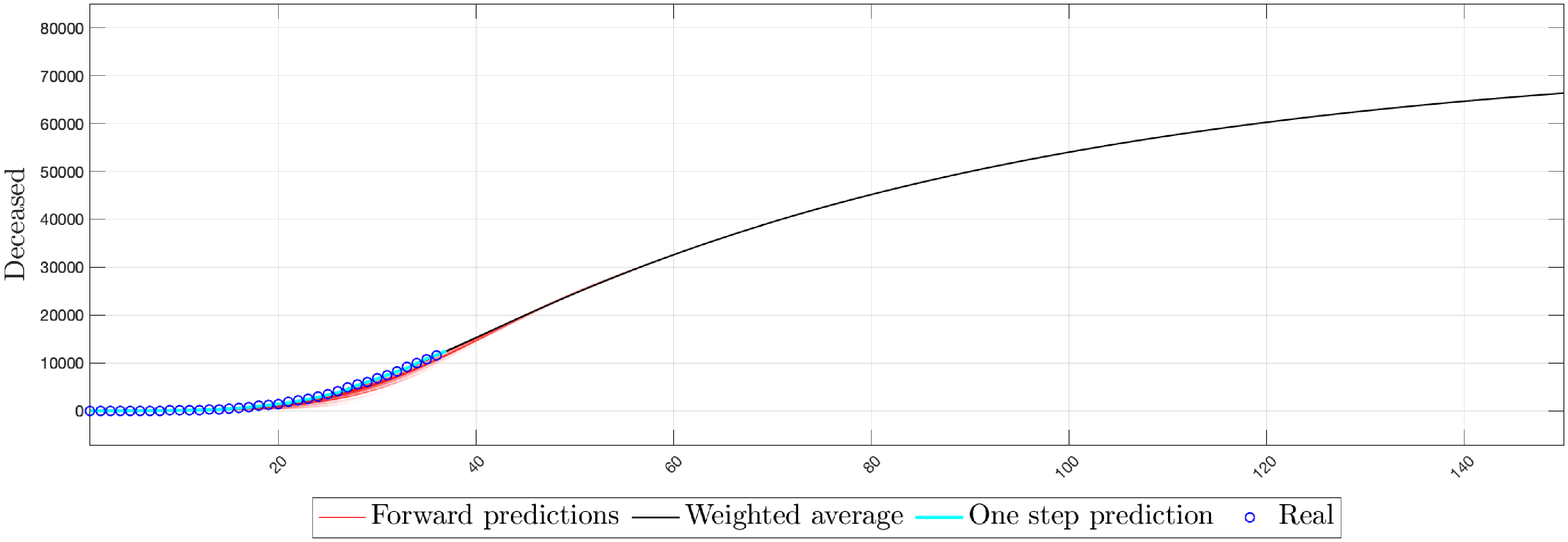}
\caption{Prediction of future values of infected, recovered, and deceased individuals in Italy using data up to March 30th, 2020.\label{fig:italy}}
\end{figure*}

\begin{figure*}[p]
\centering
\includegraphics[width=0.45\textwidth]{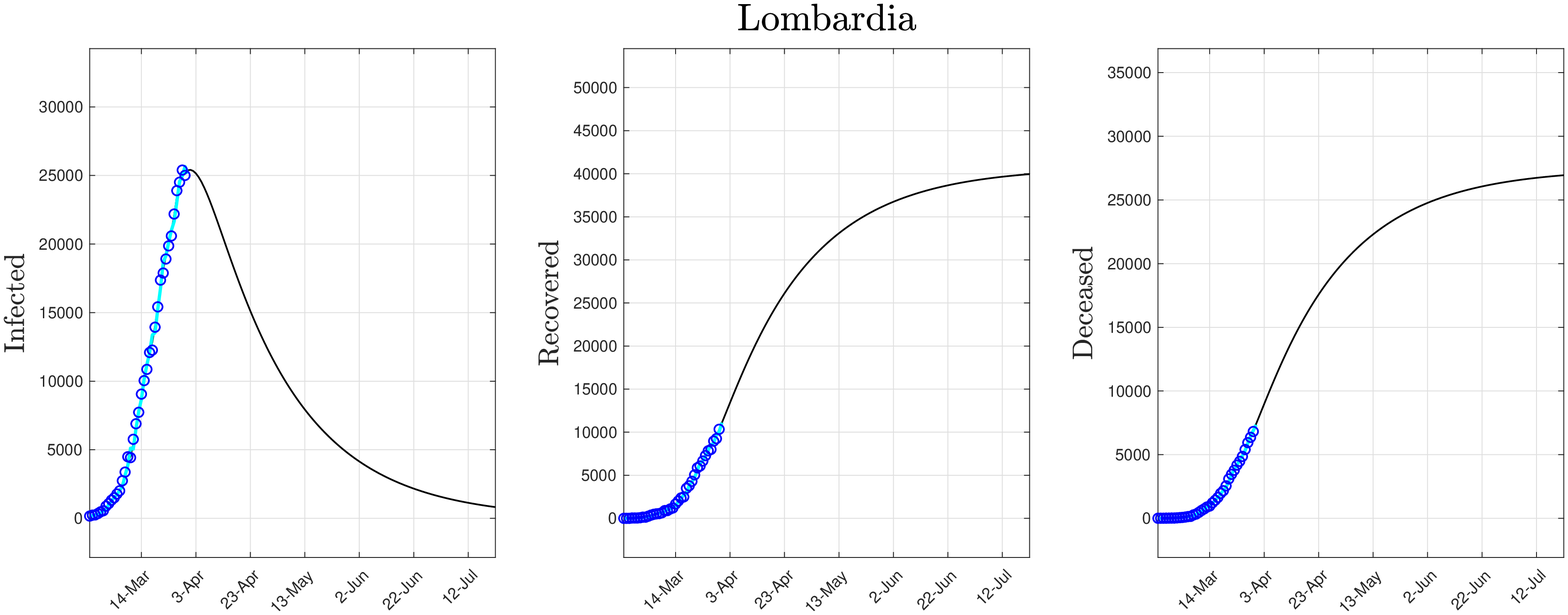}
\hfill
\includegraphics[width=0.45\textwidth]{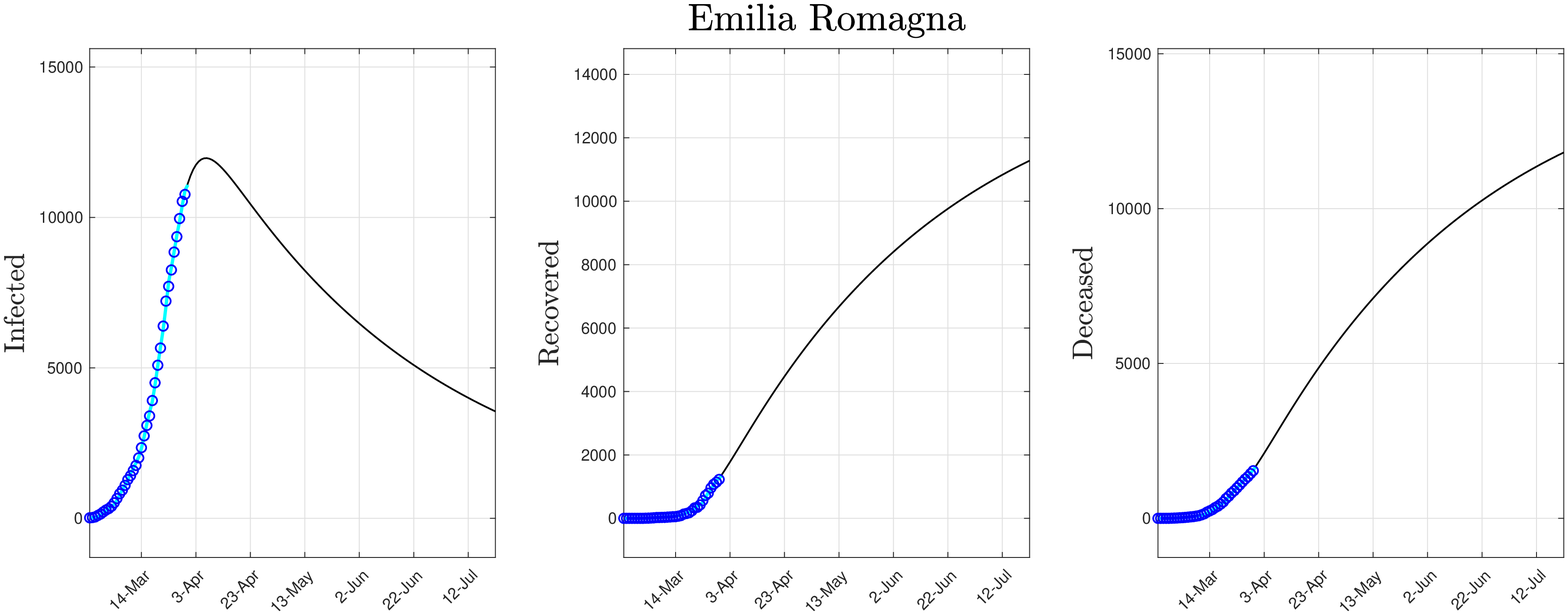}

\vspace{3ex}

\includegraphics[width=0.45\textwidth]{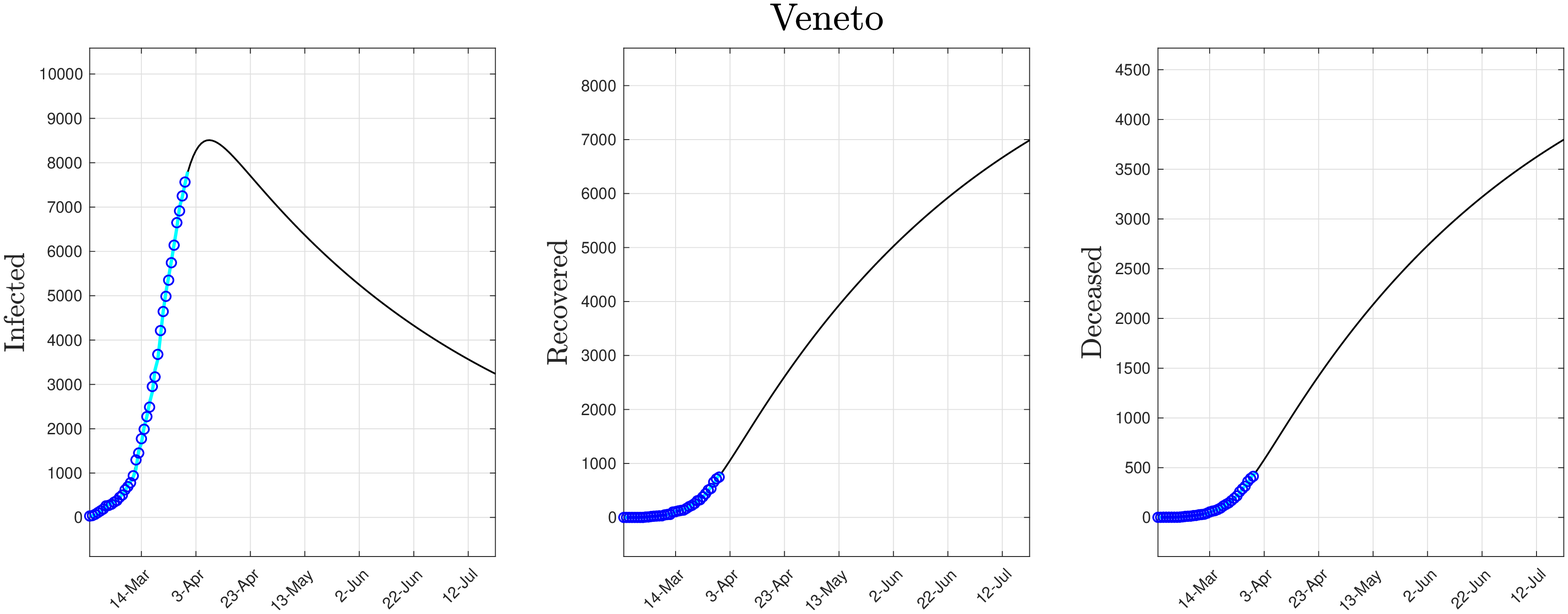}
\hfill
\includegraphics[width=0.45\textwidth]{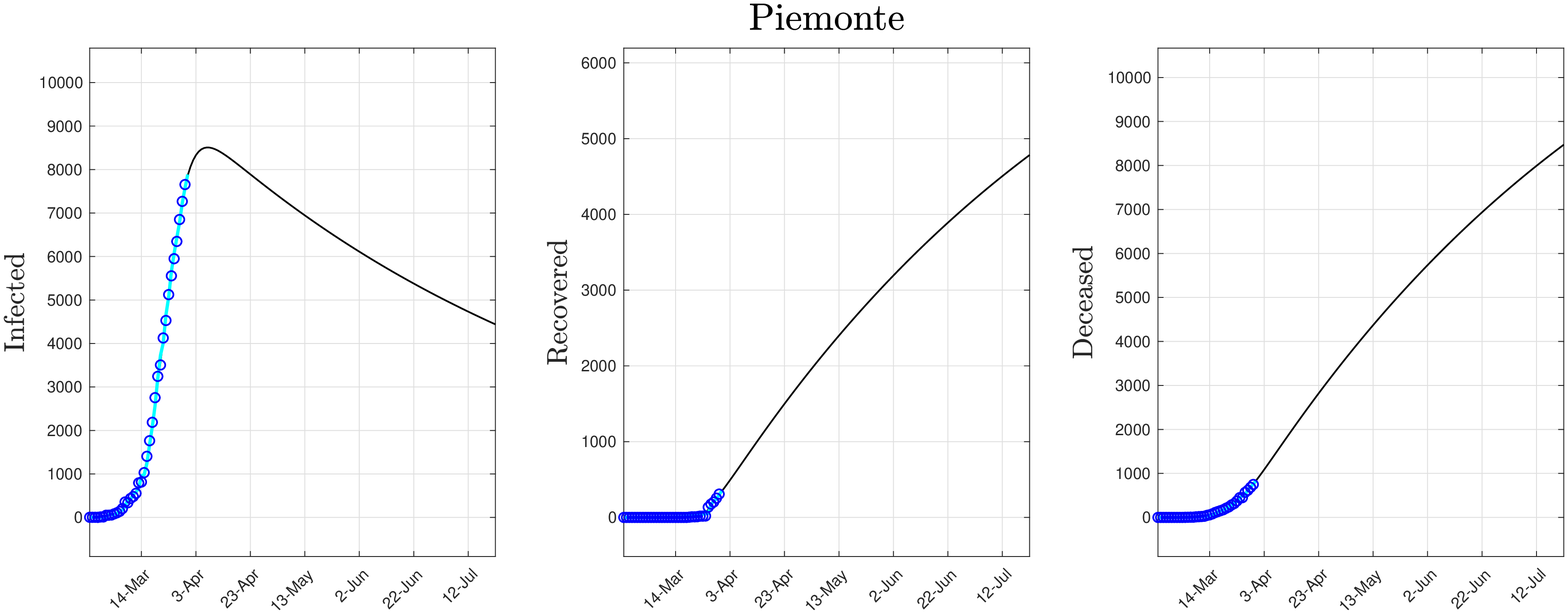}

\vspace{3ex}

\includegraphics[width=0.45\textwidth]{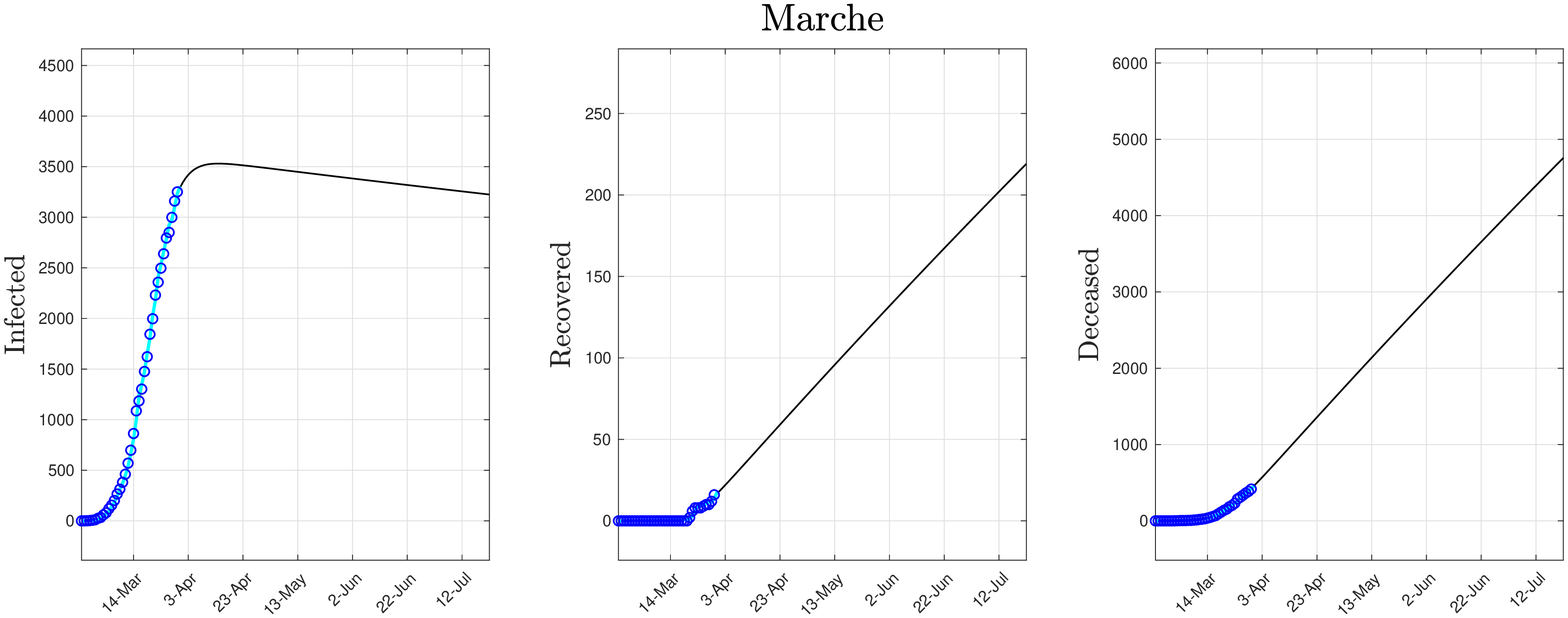}
\hfill
\includegraphics[width=0.45\textwidth]{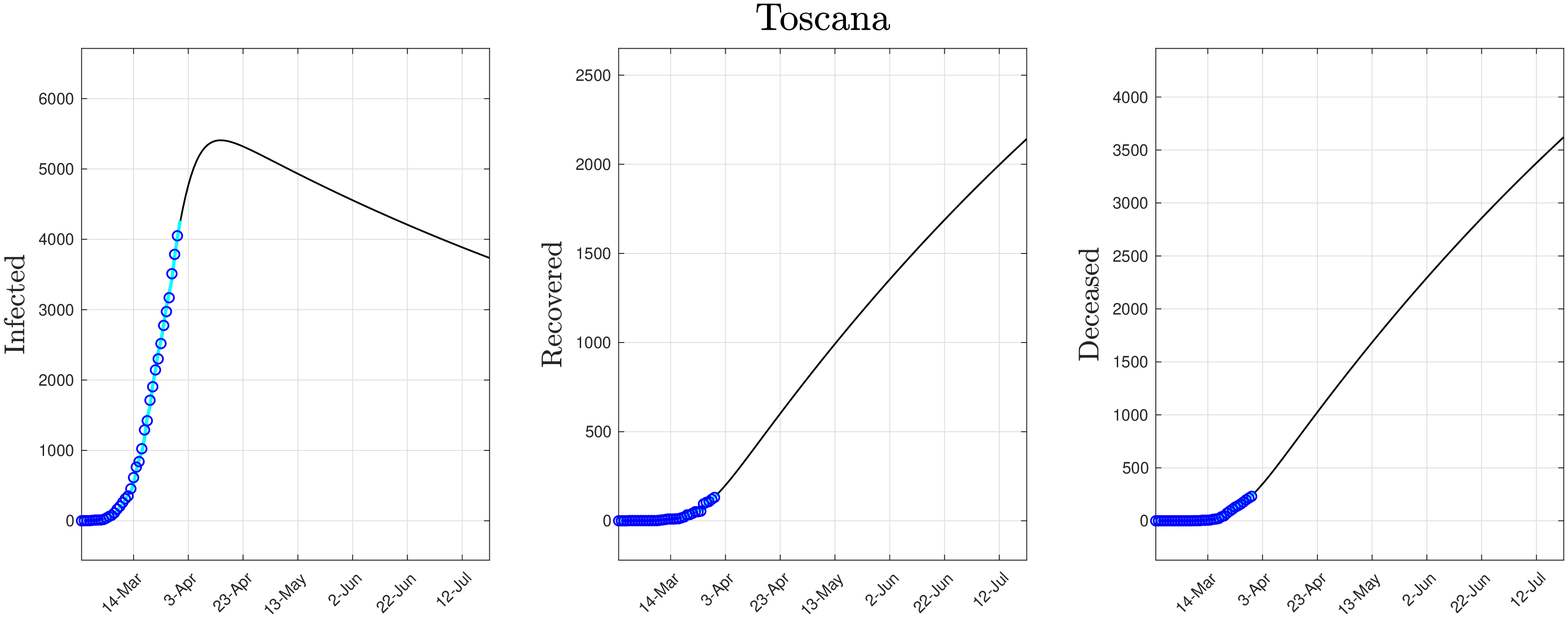}

\vspace{3ex}

\includegraphics[width=0.45\textwidth]{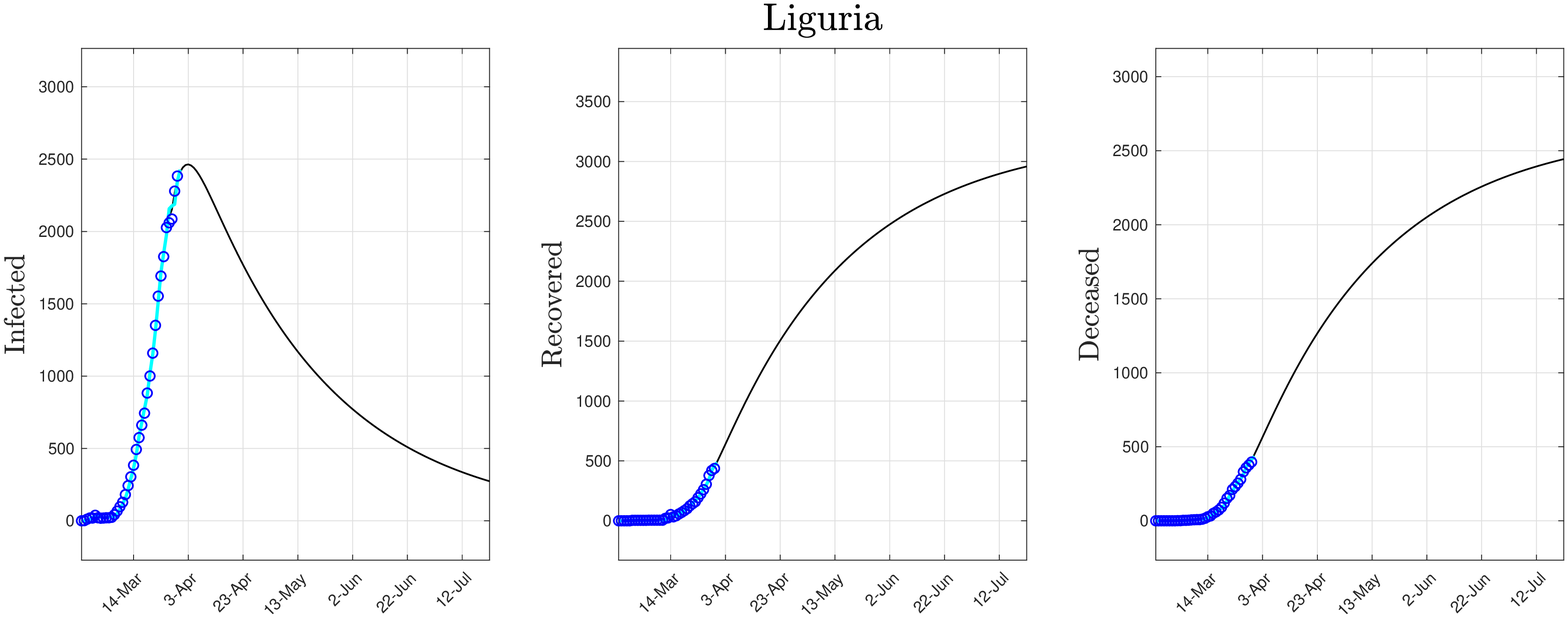}
\hfill
\includegraphics[width=0.45\textwidth]{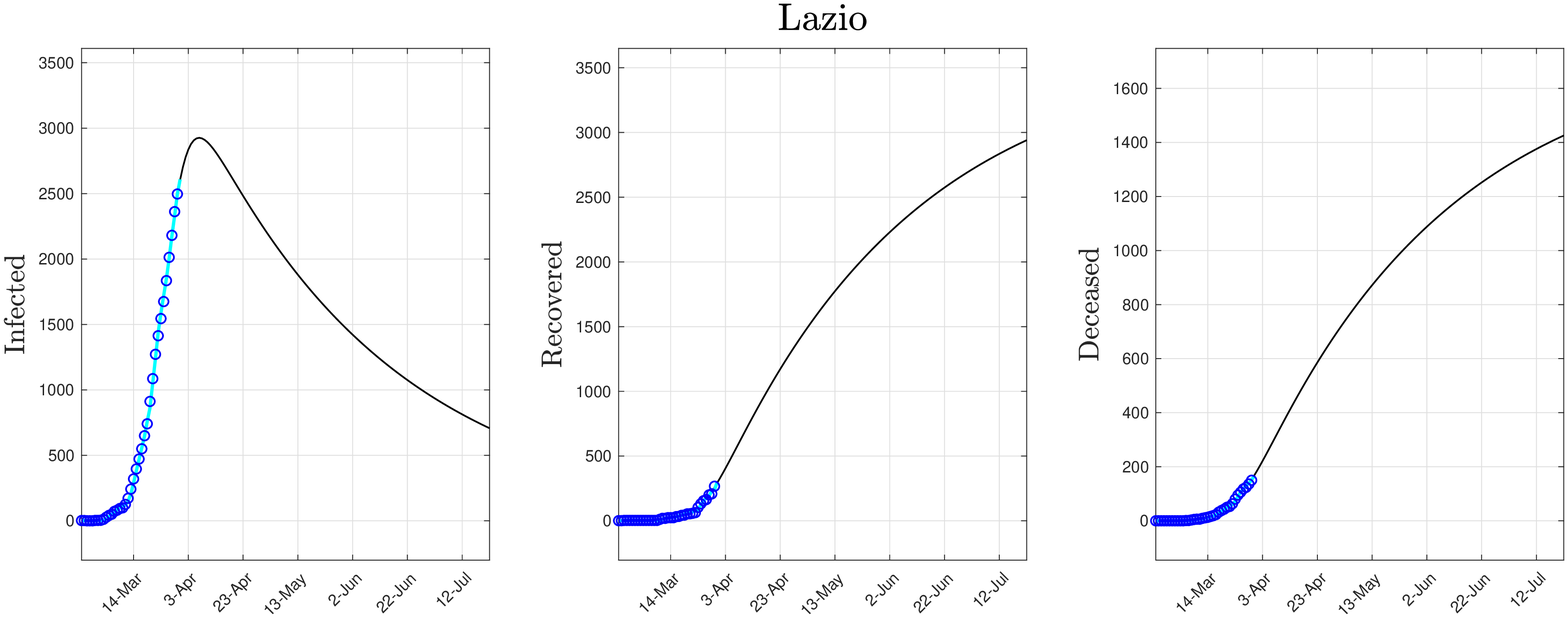}

\vspace{3ex}

\includegraphics[width=0.45\textwidth]{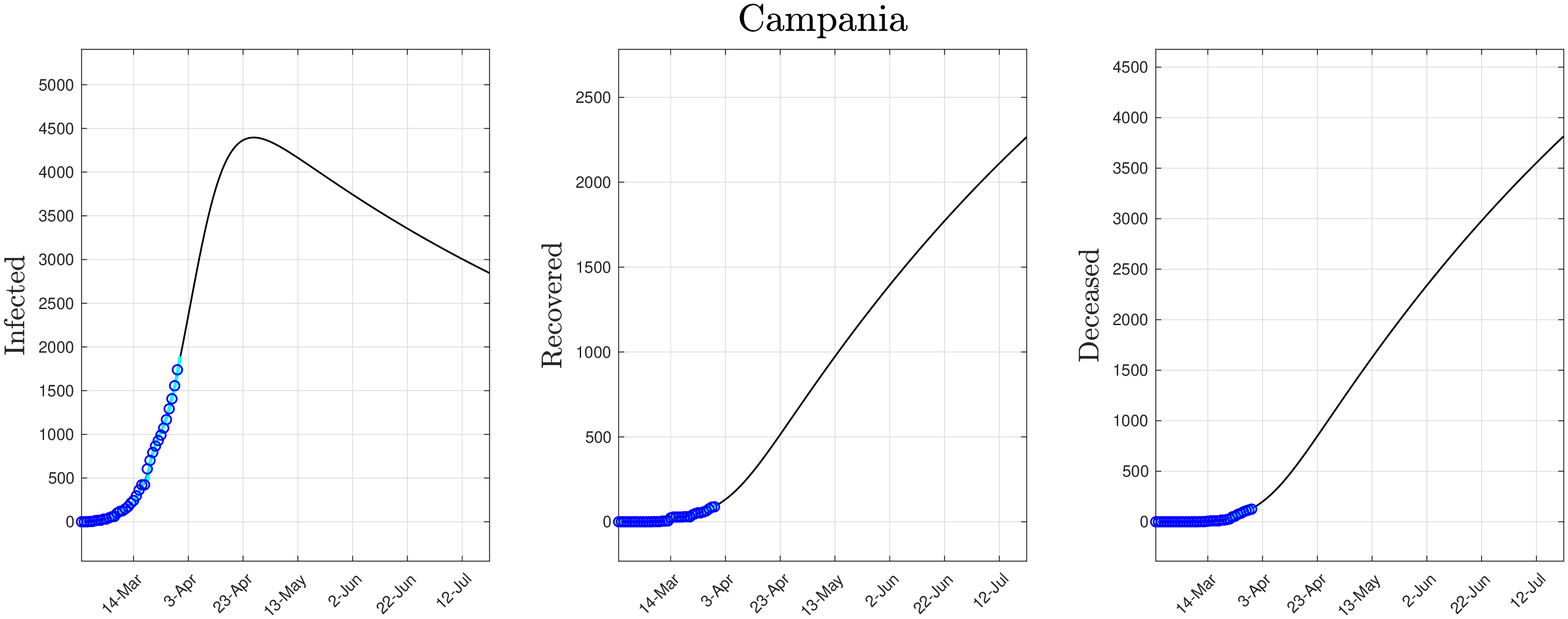}
\hfill
\includegraphics[width=0.45\textwidth]{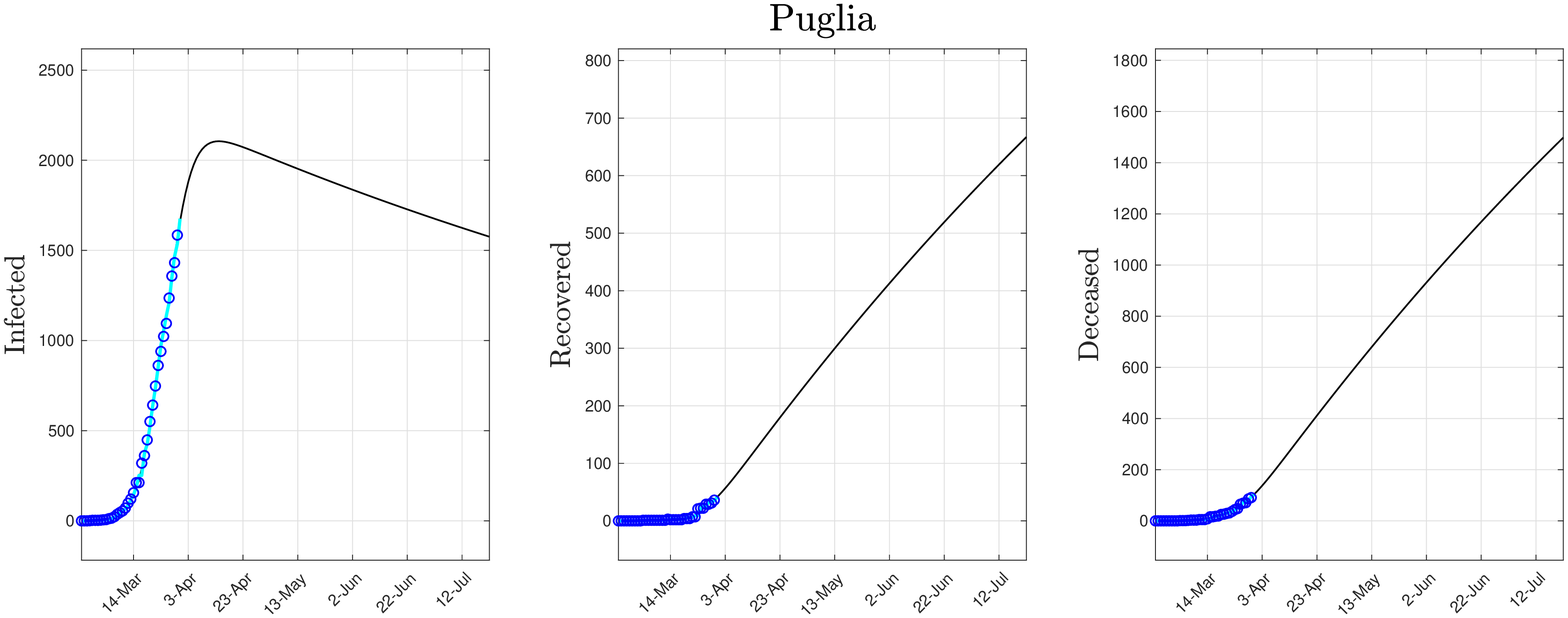}

\vspace{2ex}

\includegraphics[width=\columnwidth]{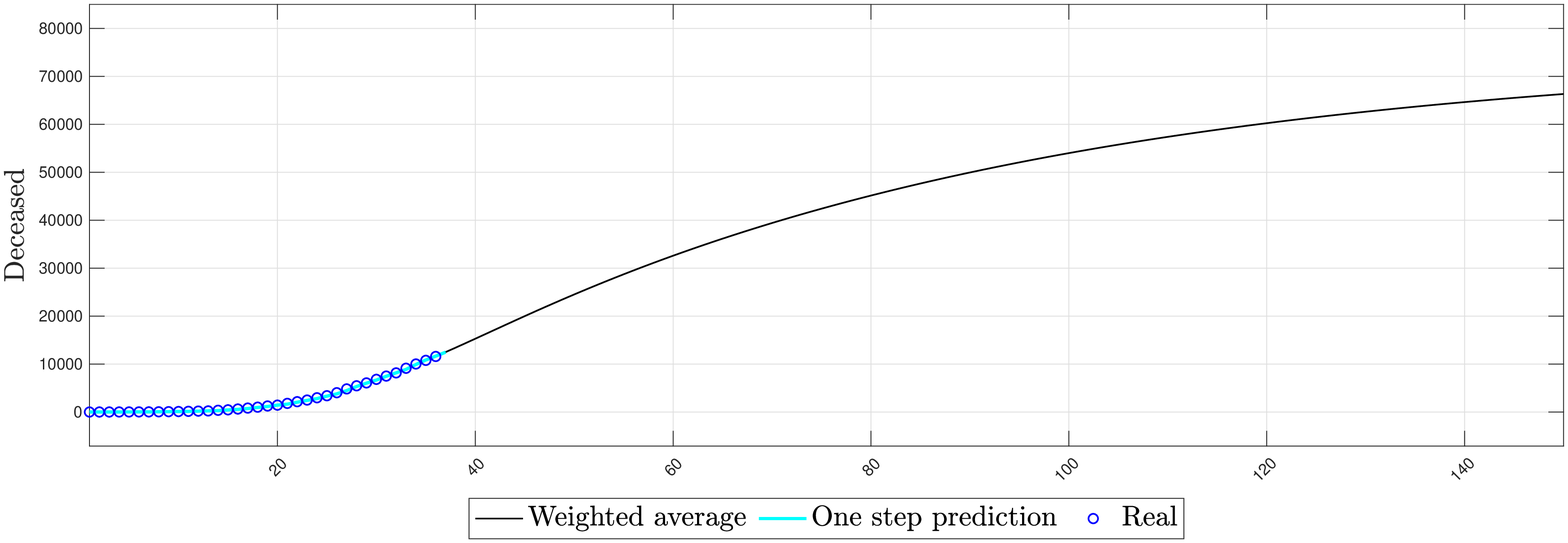}
\caption{Prediction of future values of infected, recovered, and deceased individuals in the regions of Italy 
in which the higher number of cases have been diagnosed, using data up to March 30th, 2020.\label{fig:reg}}
\end{figure*}

Once the model~\eqref{eq:SIRDm} and the initial population of susceptible individuals $S(t_0)$
have been identified, they can be used to estimate future values of detected infected $\tilde{I}(t)$, 
detected recovered $\tilde{R}(t)$,
and deceased  individuals ${D}(t)$. To this purpose, we consider  Algorithm~\ref{alg:sim}.
%
\begin{algorithm}[htb!]
\caption{Prediction of the number of $\tilde I$, $\tilde R$, and $ D$ individuals\label{alg:sim}}
\begin{algorithmic}[1]
\REQUIRE{data $\tilde{I}(t)$, $\tilde{R}(t)$, and $D(t)$, for $t=t_0,\dots,\Theta$, parameters $\alpha$, $\beta$, $\gamma$, $\tilde{\nu}$, and $\omega$,
	and total population $P$}
\ENSURE{prediction of future values of $\tilde{I}$, $\tilde{R}$, and $\tilde{D}$}
\FOR{each $t$ s.t. $\tilde{I}(t)$, $\tilde{R}(t)$, and $D(t)$ are available}
	\STATE initialize the estimates 
		\begin{align*}
			\hat{S}(t) \leftarrow \frac{\omega}{\alpha}\,P - \tilde{I}(t) - \tilde{R}(t)-\tilde{D}(t),\\
			\hat{I}(t) \leftarrow \tilde{I}(t),\quad 
			\hat{R}(t) \leftarrow \tilde{R}(t),\quad 
			\hat{D}(t) \leftarrow D(t)
		\end{align*}
		\vspace{-.5cm}
	\STATE use~\eqref{eq:SIRDm} to predict future values of $\hat{S}(\tau)$, $\hat{I}(\tau)$, $\hat{R}(\tau)$, and $\hat{D}(\tau)$
		for all $\tau \geq t$ in the prediction horizon 
	\IF{$t=t_0$}
		\STATE for all $\tau \geq t$ in the prediction horizon, let
			\begin{align*}
			\check{S}(\tau) & \leftarrow \hat{S}(\tau), &
			\check{I}(\tau) &\leftarrow \hat{I}(\tau),\\
			\check{R}(\tau) &\leftarrow \hat{R}(\tau),&
			\check{D}(\tau)&\leftarrow \hat{D}(\tau)
			\end{align*}
	\ELSE
			\STATE for all $\tau \geq t$ in the prediction horizon, let
			\begin{align*}
			\hspace{-2ex}\check{S}(\tau)&\leftarrow \tfrac{1}{2}(\check{S}(\tau)+\hat{S}(\tau)), &
			\hspace{-2ex}\check{I}(\tau)&\leftarrow \tfrac{1}{2}(\check{I}(\tau)+\hat{I}(\tau)), \\
			\hspace{-2ex}\check{R}(\tau)&\leftarrow \tfrac{1}{2}(\check{R}(\tau)+\hat{R}(\tau)), &
			\hspace{-2ex}\check{D}(\tau)&\leftarrow \tfrac{1}{2}(\check{D}(\tau)+\hat{D}(\tau))
			\end{align*}
	\ENDIF
\ENDFOR
\RETURN $\check{I}(t)$, $\check{R}(t)$, and $\check{D}(t)$
\end{algorithmic}
\end{algorithm}
This algorithm  constructs predictions $\check{I}(t)$, $\check{R}(t)$, and $\check{D}(t)$
of the future values of $\tilde{I}(t)$, $\tilde{R}(t)$, and $D(t)$, respectively, by using the model~\eqref{eq:SIRDm}
and the available data. The datum $\tilde{I}(t)$, $\tilde{R}(t)$, and $D(t)$ is used 
to compute \emph{forward predictions} $\hat{S}(\tau)$, $\hat{I}(\tau)$, $\hat{R}(\tau)$, and $\hat{D}(t)$ of the state 
variables of system~\eqref{eq:SIRDm},s for all $\tau \geq t$ in the prediction horizon.
These forward predictions are then used to update the estimates of the future values of the state variables.
In particular, letting $\hat{S}_t(\tau)$, $\hat{I}_t(\tau)$, $\hat{R}_t(\tau)$, and $\hat{D}_t(\tau)$ be the predictions at time $\tau$ obtained 
by projecting forward the datum $\tilde{I}(t)$, $\tilde{R}(t)$, and $D(t)$ available at time $t$, 
and letting $t_0+\Theta$ be the time at which the last datum $\tilde{I}(t)$, $\tilde{R}(t)$, and $D(t)$ is available,
the prediction at time $T>\Theta$ returned by Algorithm~\ref{alg:sim} is given by the \emph{weighted average}
\begin{multline*}
\check{S}(T) = \frac{1}{2^{\Theta-t_0}}\hat{S}_{t_0}(T)+\frac{1}{2^{\Theta-t_0}}\hat{S}_{t_0+1}(T)\\
+\frac{1}{2^{\Theta-t_0-1}}\hat{S}_{t_0+2}(T)+\cdots+\frac{1}{2}\hat{S}_{\Theta}(T),
\end{multline*}
whereas the prediction at time $T\leq\Theta$ is given by 
\begin{multline*}
\check{S}(T) = \frac{1}{2^{T-t_0}}\hat{S}_{t_0}(T)+\frac{1}{2^{T-t_0}}\hat{S}_{t_0+1}(T)\\
+\frac{1}{2^{T-t_0-1}}\hat{S}_{t_0+2}(T)+\cdots+\frac{1}{2}\hat{S}_{T}(T).
\end{multline*}
 Figure~\ref{fig:italy} depicts the forward predictions $\hat{S}_{t}$ (fading red lines) and their weighted average 
$\check{S}(t)$ (solid black line) obtained using Algorithm~\ref{alg:sim},
with the parameters given in Table~\ref{tab:nation}, and the one-step prediction obtained by projecting of just one step ahead the datum
available at time $t$ by using the identified model~\eqref{eq:SIRDm}.
 Algorithm~\ref{alg:sim},
has also been used for estimating the spread of COVID-19 in the most affected regions of
Italy. Figure~\ref{fig:reg} depicts the results of such predictions.

\section{Discussion}
This work has been done in the urgency of the ongoing COVID-19 contagion, with the purpose of providing a simple yet effective explanatory model for prediction of the future evolution of the contagion, and verification of the effectiveness of the containment and lockdown measures. One significant feature of the proposed model is the identification, simultaneously with model parameters, of the $\alpha$ factor that relates the number of detected positives with the unknown number of actual infected individuals in the population. For the aggregated data of Italy, such factor has been estimated to a value of about $63$. This in turns affects the real mortality rate of the infection which, if computed on the basis of the detected positives would  amount to the quite high value of $\tilde \nu= 1.18\%$, whereas if referred to the  number of actual infected individuals 
would decrease  to $\nu =\tilde \nu/\alpha= 0.019\%$.
This seemingly high proportionality factor $\alpha =63$ appears to be actually in line with today's (March 30, 2020)  estimates 
provided by Imperial College COVID-19 Response Team in  \cite{instance1290}, who foresee a total infected figure of about 5.9 million
(with an uncertainty range of [1.9 -- 15.2] million). Indeed, today's (March 30, 2020)  cumulative number of detected positive individuals in Italy
is $101739$ which, multiplied by $\alpha=63$,  yields a figure of about 6.4 million infected, that  is well within the range estimated in \cite{instance1290}.
It is to be observed that the present identification results are quite sensitive to the input data and that, due to time constraints, we could not run a suitable Monte-Carlo analysis for inferring intervals of reliability for the model parameters and predictions. Due to the large uncertainty in the data collection procedures, however, we can expect the same type of high variability reported  in \cite{instance1290}, that is, for instance,  $\pm 78\%$ uncertainty on the real number of total infected individuals.

\begin{figure}[h!tb]
\centering
\includegraphics[width=0.8\columnwidth]{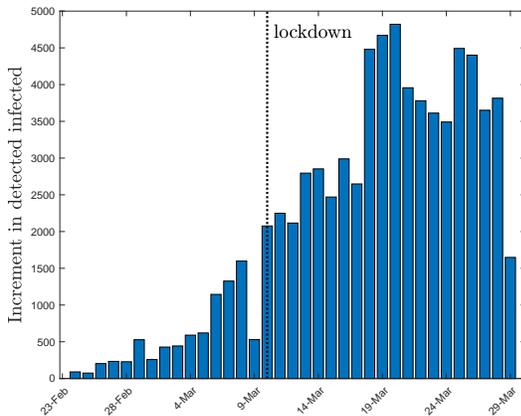}
\caption{Daily variation of the number of detected positives.\label{fig:diffpos}}
\end{figure}

Finally, notice that the data we used for tuning the model run up to March 30th, 2020. As it can be seen in Figure~\ref{fig:diffpos} most recent data show a substantial decrease of the number of infected individuals, which is imputable to the coming into effect, after a delay of about two weeks, of the lockdown measures imposed by the government. Clearly, the underlying process is non-stationary, and the predictions of the model tuned using data up to March 30th, 2020 will (hopefully) be pessimistic, as the lockdown will drastically change the underlying mechanics of the contagion.

\bibliographystyle{ieeetr}
\bibliography{biblio}

\begin{thebibliography}{10}

\bibitem{Chowell2017}
G.~Chowell, ``{Fitting dynamic models to epidemic outbreaks with quantified
  uncertainty: A primer for parameter uncertainty, identifiability, and
  forecasts},'' {\em Infectious Disease Modelling}, vol.~2, no.~3,
  pp.~379--398, 2017.

\bibitem{SIR1927}
W.~Kermack and A.~McKendrick, ``{A contribution to the mathematical theory of
  epidemics},'' {\em Proceedings of the Royal Society of London}, vol.~A 115,
  pp.~700--721, 1927.

\bibitem{Richards1959}
F.~J. Richards, ``{A flexible growth function for empirical use},'' {\em
  Journal of Experimental Botany}, vol.~10, no.~2, pp.~290--301, 1959.

\bibitem{Chowell2019}
G.~Chowell, A.~Tariq, and J.~M. Hyman, ``{A novel sub-epidemic modeling
  framework for short-term forecasting epidemic waves},'' {\em BMC Medicine},
  vol.~17, no.~1, p.~164, 2019.

\bibitem{bailey1975mathematical}
N.~T.~J. Bailey, {\em The mathematical theory of infectious diseases and its
  applications}.
\newblock New York, NY, USA: Hafner Press, 2nd~ed., 1975.

\bibitem{Brauer2017}
F.~Brauer, ``Mathematical epidemiology: Past, present, and future,'' {\em
  Infectious Disease Modelling}, vol.~2, no.~2, pp.~113--127, 2017.

\bibitem{Keeling2005}
M.~J. Keeling and K.~T. Eames, ``Networks and epidemic models,'' {\em Journal
  of the Royal Society Interface}, vol.~2, no.~4, pp.~295--307, 2005.

\bibitem{Nadini2018}
M.~{Nadini}, A.~{Rizzo}, and M.~{Porfiri}, ``Epidemic spreading in temporal and
  adaptive networks with static backbone,'' {\em IEEE Transactions on Network
  Science and Engineering}, vol.~7, no.~1, pp.~549--561, 2020.

\bibitem{Nowzari2017}
C.~Nowzari, V.~M. Preciado, and G.~J. Pappas, ``Optimal resource allocation for
  control of networked epidemic models,'' {\em IEEE Transactions on Control of
  Network Systems}, vol.~4, no.~2, pp.~159--169, 2015.

\bibitem{Pastor-Satorras2015}
R.~Pastor-Satorras, C.~Castellano, P.~Van~Mieghem, and A.~Vespignani,
  ``Epidemic processes in complex networks,'' {\em Reviews of modern physics},
  vol.~87, no.~3, p.~925, 2015.

\bibitem{PastorePiontti2014}
A.~Y. {Pastore Piontti}, M.~F. D.~C. Gomes, N.~Samay, N.~Perra, and
  A.~Vespignani, ``{The infection tree of global epidemics},'' {\em Network
  Science}, vol.~2, no.~1, pp.~132--137, 2014.

\bibitem{Pellis2015}
L.~Pellis, F.~Ball, S.~Bansal, K.~Eames, T.~House, V.~Isham, and P.~Trapman,
  ``{Eight challenges for network epidemic models},'' {\em Epidemics}, vol.~10,
  pp.~58--62, 2015.

\bibitem{Wertheim2014}
J.~O. Wertheim, A.~J. {Leigh Brown}, N.~L. Hepler, S.~R. Mehta, D.~D. Richman,
  D.~M. Smith, and S.~L. {Kosakovsky Pond}, ``{The global transmission network
  of HIV-1},'' {\em Journal of Infectious Diseases}, vol.~209, no.~2,
  pp.~304--313, 2014.

\bibitem{mizumoto2020estimating}
K.~Mizumoto, K.~Kagaya, A.~Zarebski, and G.~Chowell, ``Estimating the
  asymptomatic proportion of coronavirus disease 2019 ({COVID-19}) cases on
  board the {D}iamond {P}rincess cruise ship, {Y}okohama, {J}apan, 2020,'' {\em
  Eurosurveillance}, vol.~25, no.~10, p.~2000180, 2020.

\bibitem{gleissner1988spread}
W.~Gleissner, ``The spread of epidemics,'' {\em Applied Mathematics \&
  Computation}, vol.~27, pp.~167--171, 1988.

\bibitem{boyd2004convex}
S.~Boyd and L.~Vandenberghe, {\em Convex optimization}.
\newblock Cambridge, UK: Cambridge University Press, 2004.

\bibitem{instance1290}
S.~Flaxman, S.~Mishra, A.~Gandy, {\em et~al.}, ``{Estimating the number of
  infections and the impact of non- pharmaceutical interventions on COVID-19 in
  11 European countries},'' tech. rep., Imperial College London, 2020.

\end{thebibliography}
\end{document}